\pdfoutput=1
\documentclass[conference,a4paper,11pt]{IEEEtran}
\usepackage{mathtools}
\usepackage[dvipsnames]{xcolor}
\usepackage[keeplastbox]{flushend}
\usepackage[latin1]{inputenc}
\usepackage{times,amsmath}
\usepackage{amsfonts}
\usepackage{pstool}
\usepackage{subfigure}
\usepackage{multirow}
\usepackage{multicol}
\usepackage{enumerate}
\usepackage{graphicx}
\usepackage{mathtools, cuted}
\usepackage{MnSymbol}
\usepackage{hyperref}
\usepackage{stfloats}
\usepackage{diagbox}
\usepackage{slashbox}
\usepackage{algorithm,algorithmic}
\usepackage{bigints}
\usepackage{amsmath}
\usepackage{cite}
\begin{document}
\title{Throughput maximization of an IRS-assisted wireless powered network with interference: A deep unsupervised learning approach.}
\author{
\IEEEauthorblockN{Ahsan Mehmood$^\ast$, Omer \ Waqar$^\dagger$, and  M.\ Mahboob\ Ur\ Rahman$^\ast$  \\
$^\ast$Electrical engineering department, Information Technology University, Lahore, Pakistan \\
$^ { \dagger}$Department of Engineering, Thompson Rivers University, BC, Canada \\
$^\ast$\{msee19009, mahboob.rahman\}@itu.edu.pk,\\ $^\dagger$owaqar@tru.ca
}
}
\maketitle


\begin{abstract}
 We consider an intelligent reflecting surface (IRS)-assisted wireless powered communication network (WPCN) in which a multi antenna power beacon (PB) sends a dedicated energy signal to a wireless powered source. The source first harvests energy and then utilizing this harvested energy,  it sends an information signal to destination where an external interference may also be present. For the considered system model, we formulated an analytical problem in which the objective is to maximize the throughput by jointly optimizing the energy harvesting (EH) time and IRS phase shift matrices. The optimization problem is high dimensional non-convex, thus a good quality solution can be obtained by invoking any state-of-the-art algorithm such as Genetic algorithm (GA). It is well-known that the performance of GA is generally remarkable, however it incurs a high computational complexity. To this end, we propose a deep unsupervised learning (DUL) based approach in which a neural network (NN) is trained very efficiently as time-consuming task of labeling a data set is not required. \textcolor{black}{Numerical examples show that our proposed approach achieves a better performance-complexity trade-off as it is not only several times faster but also provides almost same or even higher throughput as compared to the GA}. 

\end{abstract}

\section{Introduction}
The 5th generation (5G) communication systems are currently being rolled out in various countries around the globe, however there is a rising concern among the practitioners that the current 5G technologies will not be able to meet extremely stringent requirements of the futuristic hyper-connected world. Moreover, as the current 5G networks comprise of massive multiple-input-multiple-output (MIMO) systems with large number of radio frequency (RF) chains, high power consumption and soaring deployment costs of these networks are also unenviable. In view of this, researchers have initiated to explore technologies for beyond 5th generation (B5G) communication systems such as 6th generation (6G) networks. B5G communication systems are expected to support massive wireless devices due to emerging use-cases of Internet of Things (IoT) \cite{9205230}. 
\textcolor{black}{However, due to scarcity in spectrum, these massive number of devices communicate in  closer or even overlapping frequency bands, which inevitably results in interference conditions\cite{8823930}. 
Along with handling co-channel interference, energizing these massive low power devices is another extremely formidable task in practice}. Although, supplying power through batteries seems to be an obvious approach, however it is not a feasible solution as frequent battery replacement is typically required. Hence, it becomes appealing to use wireless power transfer (WPT) techniques for supplying energy to these massive low-power devices\cite{9043677}.
Accordingly, a communication system that harvests energy through wireless transfer of dedicated energy signal, commonly known as wireless powered communication network (WPCN), has been proposed in \cite{7462480, 6951347}. It is envisaged that WPCN will be integral part of the B5G communication systems, however, the low energy efficiency and the uncontrollable wireless environment are key limiting factors in 
achieving the ultimate performance of the WPT techniques \cite{9122596}.

In conventional wireless communication systems, the wireless environment has been considered as uncontrollable, thus the performance optimization is limited to endpoints i.e., at user end (UE) and base station (BS). The random nature and unfavourable characteristics of the wireless environment limit the performance of any communication system including the WPCN and simultaneous information and energy transfer (SWIPT) networks. However, with the advent of the intelligent reflecting surface (IRS) recently \cite{8741198}, \cite{IRS2}, it is expected that the performance of the WPCN  and SWIPT networks can be improved significantly. 

IRS consists of an array of low-cost passive reflecting elements which can independently change the phase, amplitude or even the polarization of the incident wave impinging upon it \cite{IRS3}. Therefore, IRS possesses an ability to turn an uncontrollable wireless environment into programmable and smart entity. In other words, IRS is a break-through technology that can dramatically improve several performance metrics such as effective capacity \cite{AMAN2021101339}, outage and average error probabilities \cite{Waqar_ETT},  spectrum efficiency \cite{ref4} through configuration of phase shifts for the reflecting elements.
	


Although IRS is relatively a new concept in wireless communications, there exists numerous prior works that investigated the benefits of IRS integration in WPCN with different objective functions. For instance, joint optimization of IRS phase shift matrices corresponding to both downlink energy transfer (ET) and uplink information transfer (IT), time and power allocations for a multi-user WPCN is carried out in \cite{Lyu} with an objective to maximize the sum rate. As only passive beamforming is considered in \cite{Lyu} with a single antenna hybrid access point (HAP), multiple antennas at the HAP are taken in \cite{Zheng} and an algorithm for joint active and passive beamforming is proposed. Furthermore, a low-complexity channel estimation protocol is proposed in \cite{Mishra} that does not require any active participation from IRS. With this low-complexity channel estimation protocol, the authors design the near-optimal active beamforming at the power beacon (PB) and passive beamforming at the IRS with an aim to maximize the received power at an energy harvesting (EH) user. Moreover, the authors in \cite{9003222} consider IRS-assisted user cooperation in WPCN and derive maximum of common throughput  by jointly optimizing IRS phase shift matrices, power allocation to each user, energy and information transmission times. Reference \cite{9214497} studies the sum rate maximization problem of self-sustainable IRS-assisted multi-user WPCN in which the IRS phase shift matrices in both ET and IT phases, EH time for IRS and energy receivers, information transmission time for the information receiver are jointly optimized. More importantly, different from aforementioned works, \cite{9214497} also considers circuit power consumption of the IRS in the optimization problem and it is assumed that IRS is also wireless powered along with the energy constrained users. Moreover, unlike references mentioned above, authors in \cite{wu2021irsassisted} consider sum rate maximization problem in non-orthogonal multiple access (NOMA) WPCN by jointly optimizing IRS phase shift matrix and time allocations. Specifically, they prove that the same IRS phase shift matrix for both ET and IT phases should be used for the considered system model, which not only reduces the number of optimization variables but also lowers the feedback signaling overhead dramatically.

SWIPT is another interesting paradigm for WPT. Therefore, with an aim to enhance the performance of SWIPT networks, IRS is also incorporated into these networks in \cite{IRS_WPT, WS_PM,9133435,khalili2021multiobjective, Xu}. For instance,  IRS phase shift matrix and transmit precoding matrices are jointly optimized with the objective of meeting EH requirements and maximizing the weighted sum rate for IRS-assisted MIMO SWIPT networks in  \cite{IRS_WPT}. Similarly,  IRS phase shift matrix along with transmit precoders are jointly optimized in \cite{WS_PM} for the SWIPT network with an objective to maximize the weighted sum power at energy receivers subject to the individual signal-to-interference-plus-noise ratio (SINR) constraints at the information receivers. Furthermore,  reference \cite{9133435} considers multiple IRSs to assist the SWIPT network with the objective of minimizing transmit power of access point (AP) under the quality of service constraints namely, individual SINR constraints at all information receivers and EH constraints at all energy receivers. In particular, the phase shift matrices of all the IRSs and the transmit precoding vectors are jointly optimized with the above mentioned constraints in \cite{9133435}. In addition, very recently, \cite{khalili2021multiobjective} introduces the multi-objective optimization framework and investigates the trade-off between the sum rate maximization and the total EH maximization for the IRS-assisted SWIPT network by jointly optimizing the energy/information beamforming vectors and phase shifts of the IRS. Furthermore, in
\cite{Xu} different transmission modes for the IRS-assisted SWIPT network are realized by partitioning large IRS into small tiles. More specifically, joint optimization of the energy/information beamforming vectors and transmission modes selection corresponding to these small tiles is performed in \cite{Xu} with an objective to minimize the total transmit power at the AP under the constraints of individual SINR and EH requirements at the information and energy receivers, respectively. \textit{Although external interference is inevitable in current and future communication networks \cite{8823930}, it has been ignored in all the prior works for IRS-assisted WPCN and SWIPT networks, as evident from the aforementioned literature survey. Hence, to the best of our knowledge, this is a first work which considers an external interference for the IRS-assisted WPCN.} 

Availability of the big data and access to the low-cost graphical processing units (GPUs) for training neural networks (NNs) have increased significantly during the last few years. Owing to this, Deep Learning (DL) has been successfully employed in many fields such as computer vision, natural language processing (NLP) and wireless communications etc. Mainly, there are three mechanisms for training NNs, namely, supervised learning (SL), unsupervised learning (UL) and reinforcement learning (RL). SL is considered to be impractical, particularly for large wireless networks because labeling a data set is most often a prohibitively time-consuming task. \textcolor{black}{Moreover, RL algorithms are only suitable for problems that are formulated as Markov Decision
Processes (MDPs) \cite{Yang2021} and convergence of these algorithms is generally not easy to achieve owing to the inherent exploitation-exploration trade-off.} On the other hand, by exploiting  prior knowledge that exists in terms of the analytical frameworks of the communication theory, NNs can be efficiently and easily trained through UL because a labeled data set is not required. Motivated by the notable advantages of UL over both SL and RL, deep unsupervised learning (DUL) based solutions for various challenging non-convex optimization problems have been proposed in some recent works, e.g., \cite{Wei_Yu, Huang,Arjun,Passive_BF,Passive_BF_2} and references therein. In \cite{Wei_Yu}, DUL based power control method has been proposed for an optimization problem that maximizes the sum rate of a fading multi-user interference channel. Moreover, DUL based beamforming strategy  for the downlink MIMO systems is proposed in \cite{Huang} and it is demonstrated that the proposed strategy is much faster as compared to the conventional optimization algorithms. Similarly, DUL based solution has been proposed in \cite{Arjun} for solving computationally complex generalized assignment problems with a case-study for user-association in wireless networks. In the context of IRS-assisted communication networks, DUL based approaches have been proposed in \cite{Passive_BF} and \cite{Passive_BF_2}. In particular, \cite{Passive_BF} proposes DUL based passive beamforming design, and it is shown that the proposed approach executes much faster as compared to the non-data driven optimization method (i.e., semi-definite relaxation). The work of \cite{Passive_BF} is then extended to the multi-user scenario in \cite{Passive_BF_2} and the joint optimization of the active and passive beamformers is carried out using DUL. More importantly, it is shown in \cite{Passive_BF_2} that the execution time complexity of the proposed approach is much lower as compared to the non-data driven optimization method (i.e., block coordinate descent).

It is clear from the aforementioned discussion that DUL is an efficient learning mechanism for offline training of NNs and has orders of magnitude lesser time complexity of providing online real-time solutions when compared to the traditional non-data driven optimization algorithms. Motivated by this fact along with the observation that prior works of IRS-assisted WPCN and SWIPT networks i.e., [12]-[22] consider only conventional  non-data driven algorithms for the corresponding non-convex optimization problems, we propose a new DUL based approach in this paper. \textit{In fact, to the best of our knowledge, this is a first work that proposes a data-driven approach for the joint optimization of several parameters for the IRS-assisted WPCN}. To be more specific, our main contributions are summarized as follows:

\textbf{(i)} We first derive a closed-form expression for the end-to-end signal-to-noise and interference ratio (SINR) and then leverage this closed-form expression to formulate an optimization problem that maximizes the throughput.

\textbf{(ii)} We propose a DUL based approach for the considered optimization problem. In order to implement the proposed approach, we use a customized loss function, leverage special structural feature design and present a new NN architecture in which the output layer is split into three tensors corresponding to each optimization variable.

\textbf{(iii)} \textcolor{black}{We provide numerical examples which show that  our proposed DUL based approach achieves a better \textit{performance-complexity} trade-off when compared to the state-of-the-art benchmark optimization algorithm i.e., Genetic algorithm (GA). More specifically, we show that DUL based approach is not only many times faster (lesser time complexity) but also achieves throughput either closer or even higher than the GA. Moreover, we also demonstrate that the performance of our proposed DUL based approach is much better than the simpler ``random configuration" scheme.}

The rest of this paper is organized as follows. In Section \ref{SM}, first system model is introduced and then optimization problem is formulated. Moreover, details of our proposed DUL based approach are given in Section \ref{NN}. Numerical examples that investigate and compare performance of the proposed approach are provided in Section \ref{performance}. Finally, we conclude the paper in Section \ref{conc}.


{\textit{\textbf {Notations}}:} Scalars, vectors and matrices are represented by small, small bold and capital bold letters, respectively. 2-norm, absolute and transpose  are represented  by $||..||$, $|..|$ and $.^T$,  respectively. $vec(\mathbf{A})$ represents the vectorization of a matrix $\mathbf{A}$. $\mathbf{x}\star\mathbf{y}$ represents an element-wise operator between column vectors \textbf{x} and \textbf{y}. Furthermore, $\mathsf{ReLU}(x)\triangleq x^{+}\triangleq \text{max}(x,0)$ represents a Rectified Linear Unit activation function. $\lfloor x \rfloor$ denotes a floor function on any scalar $x$.

\section{System Model and Problem Formulation}
\label{SM}

We consider a WPCN, which consists of a dedicated PB that is equipped with $M$ antennas, an IRS with $N$ passive reflecting elements, a single antenna wireless-powered source ($\mathcal{S}$), and a single antenna destination ($\mathcal{D}$). Moreover, there exists a single antenna interferer, as shown in Fig. \ref{fig:SM}. Furthermore, block fading model is assumed which implies that all the channel coefficients remain constant for a channel coherence time ($T_c$) and then change independently following a complex circularly symmetric Gaussian (CSG) distribution with zero mean and \textcolor{black}{variance  $\sigma_{ij}^2$. Here $\sigma_{ij}^2$ $=L_cd_{ij}^{-\alpha}$ accounts for the path-loss, $d_{ij}$ denotes a distance between $i$th node and $j$th node, where $i,j$ $\in$ $\left\{\mathcal{S}, \mathcal{D}, \text{IRS, interferer}, \text{ PB and}  \text{ IRS}\right\}$, $L_c$ and $\alpha$ represent path-loss coefficient and exponent, respectively. It is worth highlighting here that following several prior works such as \cite{Passive_BF} and \cite{Zhang2021}, we consider perfect channel state information (CSI) and continuous phase shift of each reflecting element in this paper. However, investigation of an impact of discrete phase shifts \cite{Waqar_ETT} and/or imperfect CSI \cite{Yang2021} on the overall performance is an interesting but a challenging task, hence may be considered in our future work.}  A ``harvest-and-transmit" protocol \cite{Zhong} is considered i.e., during each time interval $T_c$, the energy from PB and information from $\mathcal{S}$ are transferred in two orthogonal time slots (or phases) and is described as follows:
\begin{figure}
\centering
\includegraphics[width=3.5in]{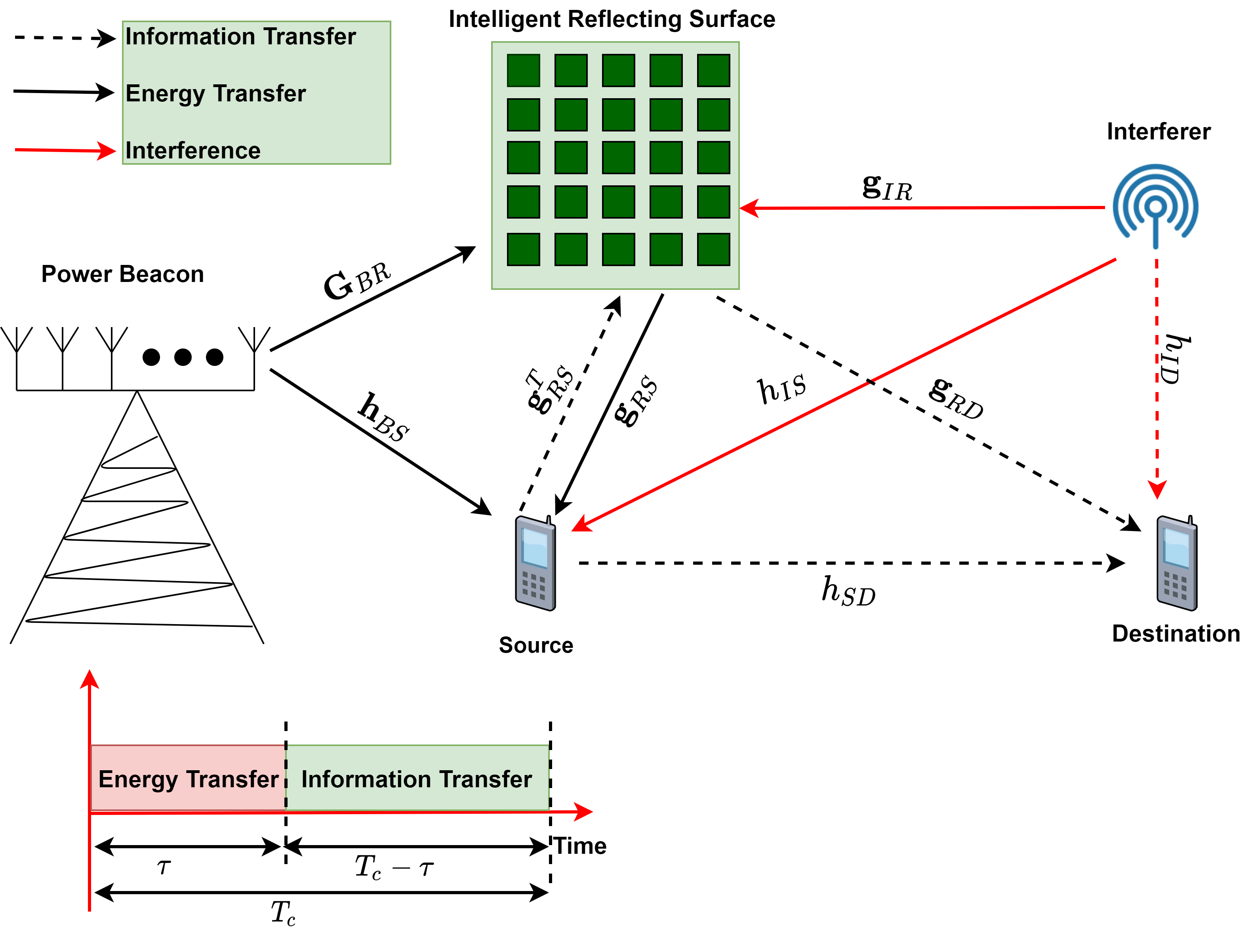}
\caption{Illustration of system model.}
\label{fig:SM}
\end{figure}


\subsubsection{Phase 1: Energy Transfer}

During a time slot of duration $\tau \in (0,1]$, an energy symbol of unit-power $\left(s_E\right)$ is transmitted by PB, which reaches at $\mathcal{S}$ through direct and IRS links. Moreover, interferer also transmits a unit-power signal ($x_I$). Thus, the signal received at $\mathcal{S}$ is the superposition of energy and interference signals transmitted through both direct and IRS channels, and can be expressed as,
\begin{multline}
     y_s  = \sqrt{P_B}[\textbf{h}^T_{BS} + \textbf{g}^T_{RS} \mathbf{\Theta}_{ET} \textbf{G}_{BR}]\textbf{w}s_E  + \\ \sqrt{P_I}[h_{IS} + \textbf{g}^T_{RS} \mathbf{\Theta}_{ET} \textbf{g}_{IR}]x_I + z_1,
\end{multline}
where $\textbf{w}\in\mathbb{C}^{M\times 1} $ denotes a unit-norm transmit beamforming vector satisfying $ \textbf{w} = \textbf{h}_{BS}+ \textbf{G}_{BR}^T \mathbf{\Theta}_{ET} \textbf{g}_{RS}/|| \textbf{h}_{BS}+ \textbf{G}_{BR}^T \mathbf{\Theta}_{ET} \textbf{g}_{RS} ||$. It is worth mentioning here that $\textbf{w}$ corresponds to the maximum ratio transmission (MRT) and is chosen  such that to maximize the EH at $\mathcal{S}$. Moreover, $P_B$ and $P_I$ represent average transmit powers of PB and interferer, respectively. Furthermore, $\textbf{h}_{BS} \in \mathbb{C}^{M\times 1} $ and $h_{IS} \in \mathbb{C}^{1\times 1}$  are direct channels from PB and interferer to $\mathcal{S}$, respectively. Similarly, $\mathbf{G}_{BR} \in \mathbb{C}^{N\times M}$, $\textbf{g}_{IR} \in \mathbb{C}^{N\times 1}$ and  $\textbf{g}_{RS} \in \mathbb{C}^{N\times  1}$  represent channels between IRS and PB or interferer or $\mathcal{S}$, respectively, as shown in Fig. 1. Furthermore, $z_1$ denotes additive white Gaussian noise (AWGN) following CSG distribution with zero mean and variance $\sigma_z^{2}$. In addition, $\mathbf{\Theta}_{ET} = \textrm{diag}\{\beta_1 e^{j\theta^1_{ET}},\beta_2 e^{j\theta^2_{ET}}, . . . , \beta_N e^{j\theta^N_{ET}} \} \in \mathbb{C}^{N\times  N}$ represents IRS phase shift matrix in ET phase with $\beta_{n} \in (0,1]$ and $\theta^n_{ET} \in [0, 2\pi)$ are the amplitude coefficient and phase shift of $n$th reflecting element, respectively.  
For ease of practical implementation and in order to maximize EH at $\mathcal{S}$, we have taken, $\beta_n = 1$ $\forall n$ and for analytical tractability we adopted linear EH model. Therefore, amount of energy harvested at $\mathcal{S}$ is given as
\begin{multline}\label{Es}
        E_s = \eta \tau T_{c} [P_B || \textbf{h}^T_{BS}+  \textbf{g}_{RS}^T \mathbf{\Theta}_{ET} \textbf{G}_{BR} ||^2 \\  + P_I |h_{IS}+\textbf{g}^T_{RS}\mathbf{\Theta}_{ET} \textbf{g}_{IR}|^2 ],
\end{multline}
where $\eta \in (0,1]$ denotes energy conversion efficiency \cite{Waqar} and its value depends on the type of EH circuitry.

\subsubsection{Phase 2: Information Transfer}
In this phase, $\mathcal{S}$ transmits an information symbol ($s_I$) with $\mathbb{E}{[|s_I|^2]}=1$ during the remaining $(1-\tau)$ time slot, as shown in Fig. 1. This information signal along with an interference signal reaches at $\mathcal{D}$ through direct and IRS paths. Therefore, the received signal at $\mathcal{D}$ can be expressed as
\begin{multline}
    y_d = \sqrt{P_S}[h_{SD}+\textbf{g}^T_{RD} \mathbf{\Theta}_{IT} \textbf{g}_{RS}]s_{I}+ \\
    \sqrt{P_I}[h_{ID}+\textbf{g}^T_{RD} \mathbf{\Theta}_{IT} \textbf{g}_{IR}]x_{I} + z_2,
\end{multline}
where  $\mathbf{\Theta}_{IT} = \mathrm{diag}\{\beta_1 e^{j\theta^1_{IT}},\beta_2 e^{j\theta^2_{IT}}, . . . , \beta_N e^{j\theta^N_{IT}} \} \in \mathbb{C}^{N \times N}$ is IRS phase shift matrix during IT phase. Similar to ET phase, we have taken $\beta_n= 1$ and $0\leq\theta^n_{IT}<2\pi$ $\forall n$. Moreover, $h_{SD}$, $h_{ID}$ and $\mathbf{g}_{RD}$ are channels from $\mathcal{S}$, interferer and IRS to $\mathcal{D}$, respectively. Furthermore, $z_2$ denotes AWGN at $\mathcal{D}$ following CSG distribution with zero mean and  variance $\sigma_z^{2}$. The transmit power of $\mathcal{S}$ is given as
\begin{equation}\label{Ps}
    P_S = \frac{E_s}{(1-\tau) T_{c}}.
\end{equation}
Moreover, SINR at $\mathcal{D}$ is given as
\begin{equation} \label{SINR}
    \gamma_D = \frac{P_S|h_{SD}+ \textbf{g}^T_{RD} \mathbf{\Theta}_{IT} \textbf{g}_{RS}|^2}{P_I|h_{ID}+\textbf{g}^T_{RD} \mathbf{\Theta}_{IT} \textbf{g}_{IR}|^2+\sigma^2_z}.
\end{equation}
Substituting  \eqref{Es} and \eqref{Ps} into \eqref{SINR}, we get \eqref{SINR_2}, as shown on top of this page.
\begin{algorithm*}[tbh]
\begin{equation}
\gamma_D =\frac{\eta \tau  
    \left[P_B || \textbf{h}^T_{BS}+  \textbf{g}_{RS}^T \mathbf{\Theta}_{ET} \textbf{G}_{BR} ||^2   + P_I |h_{IS}+\textbf{g}^T_{RS}\mathbf{\Theta}_{ET} \textbf{g}_{IR}|^2 \right]
    |h_{SD}+ \textbf{g}^T_{RD} \mathbf{\Theta}_{IT} \textbf{g}_{SR}|^2}{(1-\tau)\left[P_I|h_{ID}+\textbf{g}^T_{RD} \mathbf{\Theta}_{IT} \textbf{g}_{IR}|^2+\sigma^2_z\right]}.\label{SINR_2}
\end{equation}
\end{algorithm*}
Therefore, the throughput (ergodic capacity per unit bandwidth)  can be written as
\begin{equation} \label{capacity_eq}
    C = (1-\tau)\log_2(1+\gamma_D).
\end{equation}

It is evident from \eqref{capacity_eq} that throughput of a considered system is a function of time splitting factor ($\tau$) and phase shift matrices ($\mathbf{\Theta}_{ET}$ and $\mathbf{\Theta}_{IT}$) of IRS.  Therefore, $\tau$, $\mathbf{\Theta}_{ET}$ and $\mathbf{\Theta}_{IT}$ need to be jointly optimized in order to maximize the throughput. To this end, we formulate the following optimization problem ($\mathcal{P}$)

\begin{equation*}
\begin{aligned}
\mathcal{P:} 
& \ \ \underset{\mathbf{\Theta}_{ET},\mathbf{\Theta}_{IT}, \tau }{\text{max}}
& & C \\
&  \ \ \text{subject to}
& & 0\leq\theta_{ET}^n<2\pi, \; n = 1, \ldots, N \\
& & & 0\leq\theta_{IT}^n<2\pi, \; n = 1, \ldots, N \\
& & & 0 < \tau \leq 1. \\
\end{aligned}
\end{equation*}

It is straightforward to see that owing to coupling of several variables and unit-modulus constraints of the phase shifts,  $\mathcal{P}$ is a high dimensional non-convex optimization problem. Hence, in general, it is very difficult to find the optimal solution of $\mathcal{P}$. Nevertheless, it is worth mentioning here that a special case of $\mathcal{P}$, where for any given $\tau$,  $M=1$ (single transmit antenna) and $P_I =0$ (noise-limited scenario), $\mathcal{P}$ turns out to be a convex optimization problem which is much easier to solve. Keeping this in mind, in the sequel, we propose a new DUL based approach that offers a high-quality sub-optimal solution of $\mathcal{P}$ with much less computational overhead.

\section{ Proposed Deep Unsupervised Learning Approach}\label{NN}
\begin{figure*}
\centering
\includegraphics[width=6in]{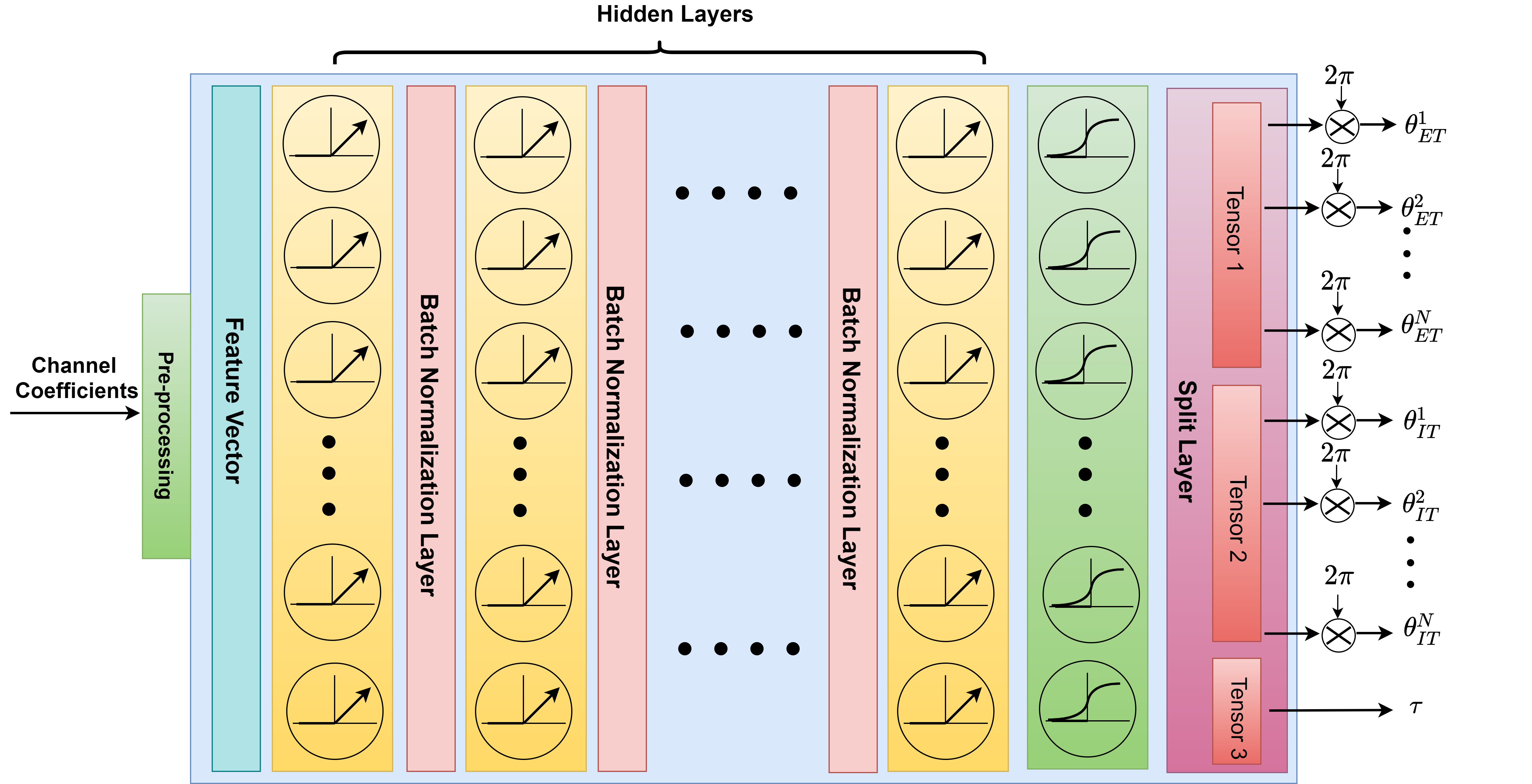}
\caption{The proposed NN architecture: IRS-Net.}
\centering
\label{fig:NN}
\end{figure*}

We leverage DL theory to find high-quality sub-optimal solution of $\mathcal{P}$. More specifically, we train the NN in an unsupervised manner. To this end, we propose an appropriate feature design, customized loss function and new NN architecture `IRS-Net' as shown in Fig. \ref{fig:NN}. Further details of our proposed approach are given in the following sub-sections.  

\begin{table*}
\caption{\label{tab:table1}Mapping between system model parameters and number of neurons of IRS-Net.}
\begin{center}
\begin{tabular}{|c||*{4}{c|}}
\hline
\backslashbox{System Model Parameters}{IRS-Net neurons}
 &\makebox[6em]{$\{n_1, n_2, . . . , n_5\}$}\\\hline\hline\textcolor{black}{
$M,N=\{2, 16 \},\{2, 32 \},\{8, 16 \}, \{4, 16\}, \{4, 32\}, \{8, 32\} ,\ P_I=0$ W}  & $\textcolor{black}{\textcolor{black}{\{2.5, 3, 2.5, 2, 0.5\}}}$\\\hline
\textcolor{black}{$M,N=\{8, 8 \},\{2, 8 \},\{4, 8 \} $, $P_I=0$ W} &\textcolor{black}{$\{3.5, 4.5, 4.5, 4,1\} $}\\\hline
\textcolor{black}{
$M= \{2, 4, 8\},\ N = \{8, 16, 32\}, P_I = 10\ \text{dBm} $  } & \textcolor{black}{$\{1.25, 2.75, 2.75, 2.05, 0.75\}$}\\\hline
\textcolor{black}{
$M= 8,\ N, P_I = \{16, 0\ \text{dBm}\}, \{8, 5\ \ \text{dBm}\}, \{8, 15\ \text{dBm}\}, \{16, 15\ \text{dBm}\}$ } & \textcolor{black}{$\{1.25, 2.75, 2.75, 2.05, 0.75\}$}\\\hline
\textcolor{black}{
$M= 8,\ N, P_I = \{8, 0\ \text{dBm}\}, \{32, 5\ \text{dBm}\}, \{16, 5\ \text{dBm}\}, \{32, 15\ \text{dBm}\}$ } & \textcolor{black}{$\{1.05, 1.35, 1.45, 1.25, 0.5\}$}\\\hline

\end{tabular}
\end{center}
 \end{table*}
\subsection{Feature Design}
In the proposed approach, first step is to pre-process all the channel coefficients and design a suitable feature vector \textbf{f}.  For this purpose, we exploit an inherent product structure of channel coefficients corresponding to each reflecting element and transmit antenna, as given in \cite{Passive_BF}. More specifically, we concatenate $vec(\textbf{G}_{BR}\star\textbf{g}_{RS})$, $\textbf{h}_{BS}$, $\textbf{g}_{IR}\star\textbf{g}_{RS}$, $\textbf{g}_{RD}\star\textbf{g}_{RS}$, $\textbf{g}_{IR}\star\textbf{g}_{RD}$, $h_{ID}$, $h_{IS}$ and $h_{SD}$ into one vector with real and imaginary components of complex numbers are concatenated separately. Hence, we have  $\mathbf{f}\in \mathbb{R}^{F_s\times 1}$, where $F_s=2 \times (NM+M+3N+3)$.
 It is worth pointing out here that for a noise-limited scenario i.e.,  in absence of interference, the size of \textbf{f} reduces to  $F_s=2 \times (MN+M+N+1)$.

\subsection{Loss Function}
We exploit the closed-form equation for throughput i.e., \eqref{capacity_eq} to define a loss function ($L$). In particular, the loss function is customized as 
\begin{equation}
L = -\frac{1}{\mathcal{|F|}} \sum_{\mathbf{f}\in \mathcal{F}} C(\mathbf{f},\theta),
\label{eq: loss_function}
\end{equation}
where $\mathcal{|F|}$ denotes the number of examples of feature vectors in a training minibatch $\mathcal{F}$ \cite{Arjun}. Moreover, $\theta$ denotes the set of trainable network parameters for IRS-Net. It is evident from (\ref{eq: loss_function}) that labeled data set is not required which reflect the unsupervised nature of IRS-Net.



\subsection{Neural Network Architecture and Training}
\textcolor{black}{Similar to \cite{Passive_BF}, our system model represents a single-user scenario, therefore a fully-connected (FC) NN is a better choice \cite{Passive_BF_2} and is adopted for our proposed IRS-Net.} The IRS-Net consists of an input layer with $F_s$  neurons, $5$ FC hidden layers and an output layer, with   $\lfloor n_1  \times F_s \rfloor$, $\lfloor n_2 \times F_s \rfloor$ ,...,  $\lfloor n_5  \times F_s \rfloor$ neurons in each hidden layer and $2N+1$ neurons in the output layer, as shown in Fig. \ref{fig:NN}. The number of neurons in each hidden layer for various system model parameters can be calculated from Table \ref{tab:table1}.  The size of each hidden layer is chosen proportional to $F_s$ so that similar architecture can be utilized even when the values of $M$ and/or $N$ change. Each hidden layer consists of $\mathsf{ReLu}$ followed by a batch normalization layer. Moreover, output layer consists of a $\mathsf{Sigmoid}$ activation function.  Here it is worth mentioning that the $\mathsf{Sigmoid}$ function inherently satisfies the unit magnitude constraints of phase shifts and constraint of time-splitting factor ($\tau$). In other words, $\mathsf{Sigmoid}$ function facilitates to meet all the constraints of $\mathcal{P}$. As shown in Fig. 2,  an output layer is followed by a split layer that partitions an output tensor of size $2N+1$ into three separate tensors, each of size $N$, $N$ and $1$. These three tensors are used to compute $\mathbf{\Theta}_{ET}$, $\mathbf{\Theta}_{IT}$ and $\tau$, respectively. It is important mentioning here that $\mathbf{\Theta}_{ET}$  and  $\mathbf{\Theta}_{IT}$ are computed simultaneously in a given  channel coherence time.  In this way, the proposed approach not only becomes computationally efficient but also its performance gets improved in terms of optimality.  

\begin{table}
\caption{\label{tab:table2}Values of other hyper-parameters for IRS-Net.}
\begin{center}
\begin{tabular}{|c||*{2}{c|}}
\hline
Hyper-parameters & Values \\\hline\hline
Training data set size & $1.2\times10^6$\\\hline
Validation data set size & $10^4$\\\hline
Test data set size & $1000$ \\\hline
Training batch size $\left(\mathcal{|F|}\right)$ & $3000$ \\\hline
Maximum number of epochs & $\textcolor{black}{500}$ \\\hline
Initial learning rate & $\textcolor{black}{10^{-3}}$ \\\hline
Optimizer & Adam \\\hline

\end{tabular}
\end{center}
 \end{table}
 



Following a common practice, Xavier initializer is used to initialize trainable network parameters. Moreover, other common hyper-parameters (their values are independent of system model parameters) are mentioned in Table. \ref{tab:table2}. We used these hyper-parameters for all the simulations unless mentioned otherwise in captions of the figures. With an aim to accelerate the convergence and achieve better performance, a staircase exponential learning rate decay is applied by using a built-in function of TensorFlow v1.15  i.e., tf.train.exponential\textunderscore decay(.) with decay\textunderscore steps = $50,000$ and decay\textunderscore rate = 0.5.  The IRS-Net is trained for maximum \textcolor{black}{500 epochs. However, in order to avoid over-fitting validation data set is used, and an early stopping is also applied in case the training loss does not decrease for consecutive 20 epochs.}

\section{Simulations and Numerical Results}
\label{performance}

\begin{algorithm*}
 \begin {equation}
 \text{Rate ratio} = \frac{\text{Throughput achievable by IRS-Net (or random scheme)}}{\text{Throughput achievable by GA}}.
 \label{rate ratio}
 \end {equation}
 \vspace{0.3cm}
 \begin {equation}
 \text{Time ratio} = \frac{\text{Computational time of  GA}}{\text{Computational time of IRS-Net}}.
 \label{time ratio}
 \end {equation}
 \end {algorithm*}

 \begin{table*}
\caption{\label{tab:table2_1} Machine specifications.}
\begin{center}
\begin{tabular}{|c||*{2}{c|}}
\hline

Specification & Value \\\hline\hline
Central Processing Unit (CPU)& Intel Core i7-4770 @ 3.40 GHz\\\hline
Random Access Memory (RAM) & $4$ GB\\\hline
Operating system & $64$bit Microsoft Windows \\\hline

\end{tabular}
\end{center}
 \end{table*}

\textcolor{black}{In this section, the numerical results are presented to validate the performance of proposed IRS-Net. The system parameters that are used to generate results are as follows: $d_{RS}=d_{IS}=d_{RD}=d_{IR}= 15$m, $d_{SD}=d_{BS}=25$m, $d_{ID}= 30$m and $L_c = 10^{-3}$. Furthermore,  $\alpha$ is equal to $2.2$ for IRS-related links and its value is equal to $2.57$ for all other links.  Moreover, $P_B = 10$ W, $\sigma^2_z = -104$ dBm, $\eta = 1$, $P_I = 0$ W for noise-limited scenario and $P_I = \left\{0, 5, 10, 15\right\}$ dBm for interference scenario. Moreover, to compare the  performance of proposed IRS-Net  with that of the benchmark optimization algorithm i.e., GA and in order to have fair comparison, \textcolor{black}{ the proposed IRS-Net and GA } utilized the same test data set to compute the throughput (or Ergodic capacity). More specifically,  GA (and trained IRS-Net) jointly optimizes the IRS phase shift matrices and $\tau$ for each feature vector and then throughput is computed by averaging over $1000$ feature vectors, i.e., over the entire test data set. Moreover, another baseline scheme ``random configuration" albeit gives performance far from optimal but much simpler than GA, is also considered. In this baseline scheme, phase shift matrices are first generated randomly using uniform distribution, then for these given random phase shift matrices, $\tau$ is generated via GA. TensorFlow v1.15 and Python v3.7.4 are used to obtain results using the IRS-Net and GA, respectively. All simulation results are generated on the same machine with the specifications given in Table. 3. GA is run either for 5 or 20 iterations in all examples.}
 Furthermore, in order to measure the performance of the IRS-Net, we define two new performance metrics, namely, `rate ratio' and `time ratio'. \textcolor{black}{The mathematical forms of these two metrics are given in \eqref{rate ratio} and \eqref{time ratio}}.  Hence, considering these definitions, we aim for high values of both `rate ratio' and `time ratio' metrics. 

\begin{figure}
\centering
\includegraphics[width=3.5in]{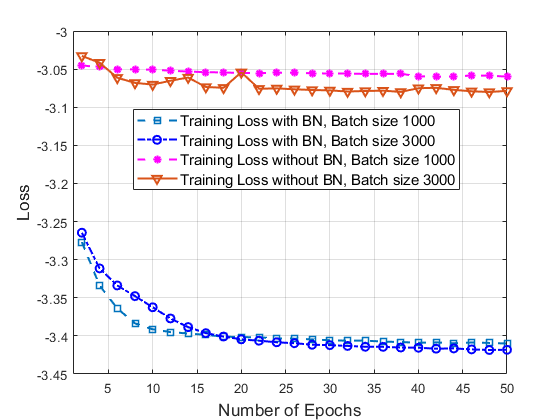}
\caption{Impact of batch normalization (BN) and batch size on training and validation losses. $P_B=10$ W, $\eta=1$, $P_I=0$ W, $M=2$ and $N=32$.}
\centering
\label{fig:Effect_BN}
\end{figure}

Through extensive simulations, it is also observed that the batch normalization is an important layer for accelerated training of the IRS-Net. In particular, \textcolor{black}{it is noticed from Fig. \ref{fig:Effect_BN} that without batch normalization layers, the loss function does not decrease. Nevertheless, it is also seen that learning is significantly improved by adding the batch normalization layers,  Thus, batch normalization is a very important layer and is adopted in IRS-Net, as shown in Fig. 2}. It certainly accelerates training (or learning) of IRS-Net and ultimately increases the overall performance of the proposed approach. A similar impact of batch normalization  is also observed in \cite{Passive_BF}. As evident from Fig. \ref{fig:Effect_BN} that training loss function converges and achieves smallest value for a batch size of 3000, this value of batch size is taken for the subsequent simulation results.

\begin{figure} 
\centering
\includegraphics[width=3.6in]{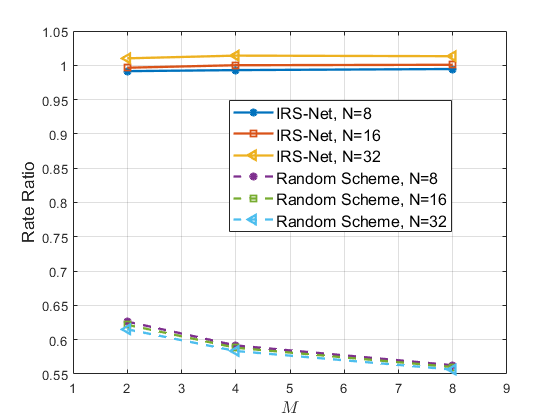}
\caption{\textcolor{black}{Impact of $M$ and $N$ on the rate ratio  in noise-limited scenario. GA: 5 iterations}}
\centering
\label{fig:rate_m}
\end{figure}

\begin{figure} 
\centering
\includegraphics[width=3.6in]{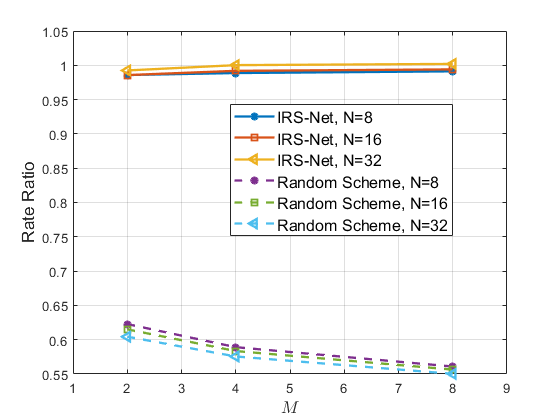}
\caption{\textcolor{black}{Impact of $M$ and $N$ on the rate ratio  in noise-limited scenario. GA: 20 iterations}}
\centering
\label{fig:rate_m_1}
\end{figure}

\subsection{Noise-limited scenario}
 A noise-limited scenario is first considered i.e., $P_I=0$ W. In this scenario, an impact of the number of transmit antennas and/or reflecting elements on the `rate ratio' is analyzed by executing GA for 5 and 20 iterations   in Fig. \ref{fig:rate_m} and Fig. \ref{fig:rate_m_1}, respectively. It is clear from Fig. \ref{fig:rate_m} and Fig.\ref{fig:rate_m_1}  that as expected, the performance of random scheme is far from that achievable by the GA because its rate ratio is quite low (much smaller than unity). \textcolor{black}{ On the other hand, the rate ratios for IRS-net are either very close or even greater than unity. The value of rate ratio greater than unity implies that the throughput achievable by the IRS-Net is higher than that of the GA. Interestingly, in comparison with the GA, rate ratio of IRS-Net increases with increasing the number of transmit antennas as well as number reflecting elements. For instance, the throughput of IRS-Net surpasses that of the GA when $M = 8$ and $N = 32$, as shown in Fig. \ref{fig:rate_m}, Fig. \ref{fig:rate_m_1} and Table. 4. }
 
 \begin{figure} 
\centering
\includegraphics[width=3.4in]{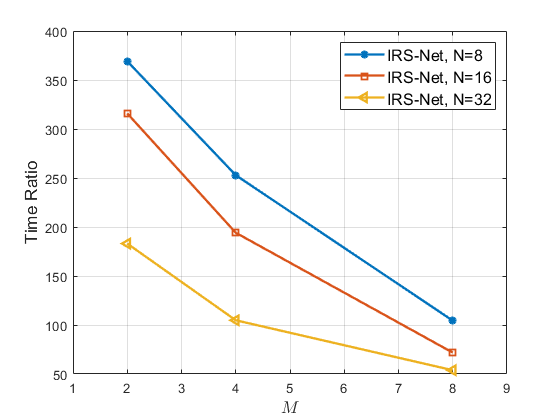}
\caption{\textcolor{black}{Impact of $M$ and $N$ on time ratio for the IRS-Net in noise-limited scenario. GA: 5 iterations}}
\centering
\label{fig:TR_m}
\end{figure}

\begin{figure} 
\centering
\includegraphics[width=3.6in]{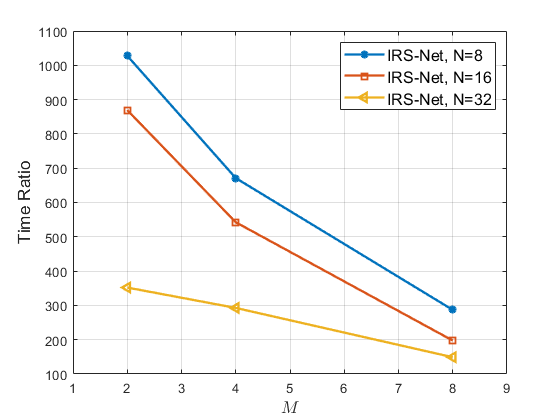}
    \caption{\textcolor{black}{Impact of $M$ and $N$ on time ratio for the IRS-Net in noise-limited scenario. GA: 20 iterations}}
\centering
\label{fig:TR_m_1}
\end{figure}

\begin{table*}
\caption{\label{tab:table3} Computational times and throughputs achievable by GA \textcolor{black}{(in 5 iterations)}, IRS-Net and Random Configuration Scheme. Throughputs for IRS-Net and Random Configuration Scheme are expressed in terms of rate ratios using eq. (9). }

\begin{center}
\begin{tabular}{|l||*{5}{c|}}
\hline
\backslashbox{Parameters}{Algorithms}
&\makebox[3em]{\textcolor{black}{GA } }&\makebox[8em]{\textcolor{black}{IRS-Net (rate ratio) }}&\makebox[11em]{\textcolor{black}{Random Scheme (rate ratio)}}&\makebox[3em]{\textcolor{black}{GA (Time)} }&\makebox[6em]{\textcolor{black}{IRS-Net (Time)}} \\\hline\hline
\textcolor{black}{$M=2, N=8$ } & \textcolor{black}{3.2871 bits/sec/Hz}&\textcolor{black}{0.991}&\textcolor{black}{0.626}&\textcolor{black}{203.03 ms }&\textcolor{black}{0.55 ms} \\\hline
\textcolor{black}{$M=4, N=16$ }  &\textcolor{black}{4.0975 bits/sec/Hz}&\textcolor{black}{1.000}&\textcolor{black}{0.588}&\textcolor{black}{219.38 ms }&\textcolor{black}{1.13 ms} \\\hline
\textcolor{black}{$M=8, N=32$}  &\textcolor{black}{4.9177 bits/sec/Hz}&\textcolor{black}{1.013}&\textcolor{black}{0.556 }&\textcolor{black}{270.36 ms}&\textcolor{black}{5.02 ms}\\\hline
\end{tabular}

\end{center}
\end{table*}

\textcolor{black}{Similarly, the effect of the number of transmit antennas and reflecting elements on the `time ratio' is depicted in Fig. \ref{fig:TR_m} and Fig. \ref{fig:TR_m_1}. Interestingly, the trend for `time ratio' is opposite to that of the `rate ratio' i.e., `time ratio' decreases with increasing $M$ and/or $N$. Moreover, both Fig. \ref{fig:TR_m} and Fig. \ref{fig:TR_m_1} show that IRS-Net is up to  $369.1$ and $1028$ times faster when compared to the GA that runs for 5 and 20 iterations, respectively.  Thus, it is concluded from aforementioned discussion that for $M=8$ and $N = 32$, our proposed IRS-Net is almost $54$ times more computational efficient and  achieves higher throughput (rate ratio = $1.013$) as compared to the GA that runs for 5 iterations. Similarly, as compared to the GA that runs for 20 iterations, IRS-Net obtains slightly higher throughput (rate ratio = $1.002$) with $149$ times faster speed. This shows that the proposed IRS-Net achieves a better performance-complexity trade-off in general, however the number of transmit antennas and the reflecting elements have an impact on this trade-off. }

\subsection{Interference scenario}
\begin{figure}
\centering
\includegraphics[width=3.6in]{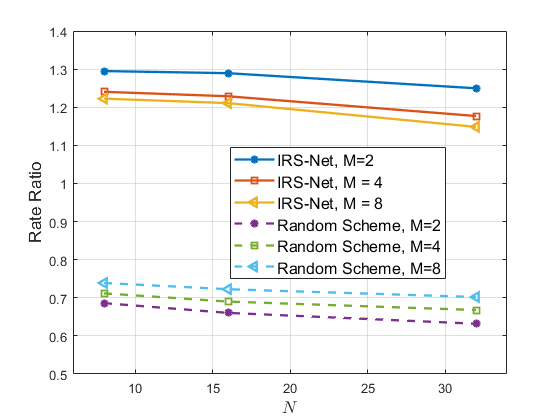}
\caption{\textcolor{black}{Impact of $M$ and $N$ on the rate ratio in presence of interference with $P_I=10$ dBm. GA: 5 iterations.}}
\centering
\label{fig:RA_Pi5_N}
\end{figure}

\begin{figure}
\centering
\includegraphics[width=3.6in]{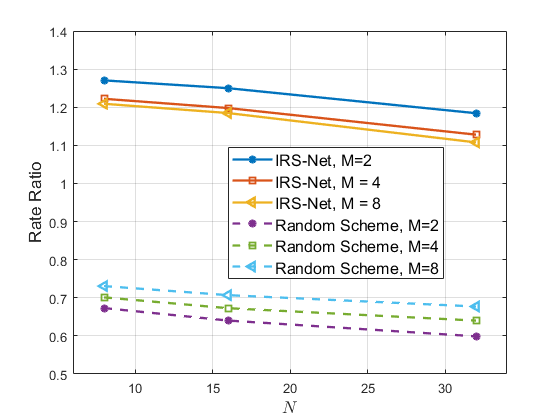}
\caption{\textcolor{black}{Impact of $M$ and $N$ on the rate ratio in presence of interference with $P_I=10$ dBm. GA: 20 iterations.}}
\centering
\label{fig:RA_Pi5_N_1}
\end{figure}

\begin{figure}
\centering
\includegraphics[width=3.4in]{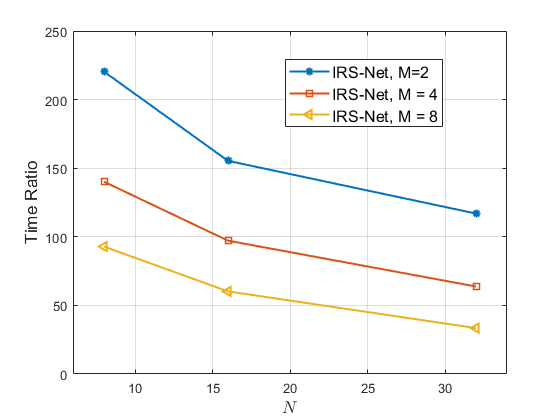}
\caption{\textcolor{black}{Impact of $M$ and $N$ on time ratio for the IRS-Net in presence of interference with $P_I=10$ dBm. GA: 5 iterations.}}
\centering
\label{fig:TR_Pi5_N}
\end{figure}

\begin{figure}
\centering
\includegraphics[width=3.4in]{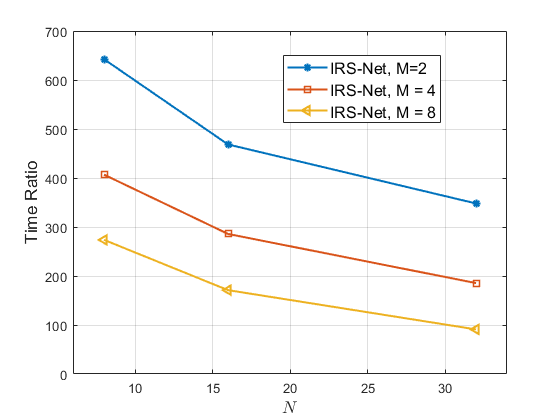}
\caption{\textcolor{black}{Impact of $M$ and $N$ on time ratio for the IRS-Net in presence of interference with $P_I=10$ dBm. GA: 20 iterations.}}
\centering
\label{fig:TR_Pi5_N_1}
\end{figure}

Now, performance of IRS-Net is analyzed in the presence of interference. \textcolor{black}{The impact of the number of transmit antennas and reflecting elements on the rate ratios is shown in  Fig. \ref{fig:RA_Pi5_N} and Fig. \ref{fig:RA_Pi5_N_1} when GA runs for 5 and 20 iterations, respectively. Contrary to the noise-limited scenario, it is evident from Fig. \ref{fig:RA_Pi5_N} and Fig. \ref{fig:RA_Pi5_N_1} that `rate ratio' corresponding to both IRS-Net decreases with increasing $M$ and/or $N$. Moreover, the trend for `time ratio' is similar to that of the noise-limited scenario and is demonstrated in Fig. \ref{fig:TR_Pi5_N}  and Fig. \ref{fig:TR_Pi5_N_1}. It is clear from Fig. \ref{fig:RA_Pi5_N} to Fig. \ref{fig:TR_Pi5_N_1}, that in the presence of interference and for several values of $M$ and $N$, IRS-net achieves much better performance-complexity trade-off as compared to the random scheme and the GA. As an example, for $M=2$ and $N=8$, IRS-Net is not only $220$ and $642$  times faster but also the achievable throughputs are higher (i.e., rate ratios are $1.29$ and $1.27$) when compared to the GA for 5 and 20 iterations, respectively.}

\begin{figure}
\centering
\includegraphics[width=3.5in]{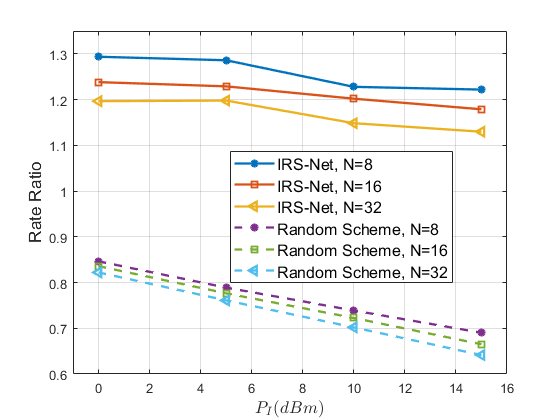}
\caption{\textcolor{black}{Impact of $P_I$ on rate-ratio in the presence of interference with $M=8$. GA: 5 iterations.}}
\centering
\label{fig:impact_Pi_11}
\end{figure}

Furthermore, an effect of the strength of interference power on the rate ratio \textcolor{black}{is analyzed in Fig. \ref{fig:impact_Pi_11}}. It is observed that the `rate ratios' for the IRS-Net and random configuration scheme decrease as interference power increases. \textcolor{black}{However, it is quite notable that the achievable throughput of IRS-Net is much higher than that of the GA as the corresponding rate ratios are greater than unity for various interference power levels. }

\begin{figure*}
     \centering
     \begin{subfigure}
         \centering
         \includegraphics[width=3.4in]{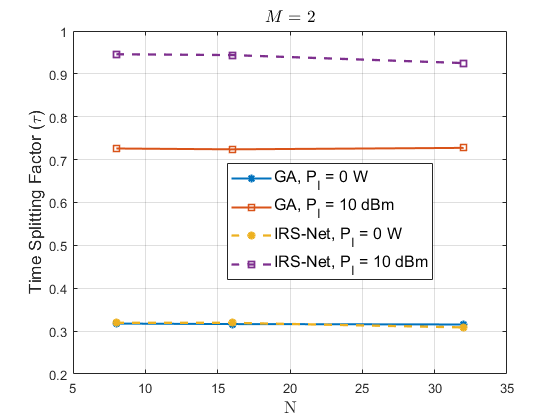}
         \label{fig:y equals x}
     \end{subfigure}
     \hfill
     \begin{subfigure}
         \centering
         \includegraphics[width=3.4in]{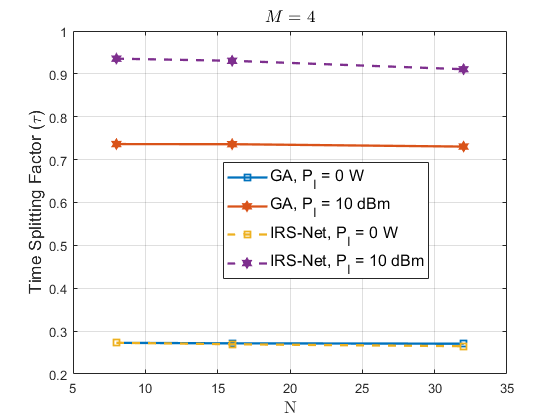}
         \label{fig:three sin x}
     \end{subfigure}
     \begin{subfigure}
         \centering
         \includegraphics[width=3.4in]{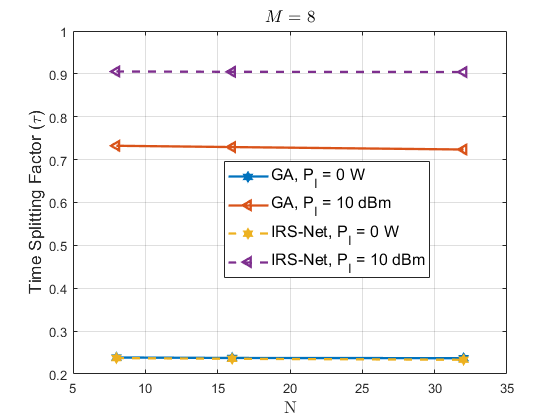}
         \label{fig:five over x}
     \end{subfigure}
        \caption{Impact of $M$ and $N$ on time-splitting factor ($\tau$). GA: 5 iterations.}
        \label{fig: Impact_on_tau}
\end{figure*}

Next the impact of the number of transmit antennas and the reflecting elements on $\tau$ is demonstrated in Fig. \ref{fig: Impact_on_tau}. It is evident from Fig. \ref{fig: Impact_on_tau} that $\tau$ decreases slightly with increasing values of $N$ and/or $M$. This is due to the fact that strength of energy beam gets improved with increasing a value of $N$ and/or $M$. However, significantly more time is required for energy transfer (i.e., higher $\tau$) when there exists an interference. This is due to the fact that larger transmit power is required at $\mathcal{S}$ in order to compensate the loss in SINR and to achieve higher throughput.

\section{Conclusion}
\label{conc}
In this paper, we considered a wireless powered communication network with IRS and external interference. We formulated an optimization problem for the considered system that involves joint optimization of phase shift matrices of IRS and time-splitting factor of energy harvesting protocol. We showed that the formulated problem is non-convex in general, therefore, it is not feasible to obtain an optimal solution within a short time period e.g.,  \textit{channel coherence time}. \textcolor{black}{To this end, we proposed a deep unsupervised learning based approach that has a better performance-complexity trade-off as compared to the GA and the random configuration scheme. Importantly, we showed that our proposed scheme provides a throughput either very close or even higher than that of the GA in a much shorter time.} Hence, considering the facts that channel coherence interval is generally very small and obtaining  a labled data set for training a NN is a time-consuming task, the utility of our  deep unsupervised learning approach is noteworthy as it offers a practically viable and high quality solution for the considered non-convex and challenging problem.

\footnotesize{
\bibliographystyle{IEEEtran}
\bibliography{main}
}
\vfill\break
\end{document}